\documentclass[twocolumn]{jpsj2}
\def\vec#1{\mbox{\boldmath $#1$}}

\title{Fourth Order Perturbation Theory for Normal Selfenergy\\ in Repulsive Hubbard Model}

\author{%
Shogo SHINKAI\thanks{E-mail address: shogo@scphys.kyoto-u.ac.jp} , Hiroaki IKEDA and Kosaku YAMADA
}

\inst{%
Depertment of Physics, Kyoto University, Sakyo-ku, Kyoto 606-8502
}

\recdate{\today}

\abst{We investigate the normal selfenergy and the mass enhancement factor in the Hubbard model on the two-dimensional square lattice. Our purpose in this paper is to evaluate the mass enhancement factor more quantitatively than the conventional third order perturbation theory. We calculate it by expanding perturbatively up to the fourth order with respect to the on-site repulsion $U$. We consider the cases that the system is near the half-filling, which are similar situations to high-$T_c$ cuprates. As results of the calculations, we obtain the large mass enhancement on the Fermi surface by introducing the fourth order terms. This is mainly originated from the fourth order particle-hole and particle-particle diagrams. Although the other fourth order terms have effect of reducing the effective mass, this effect does not cancel out the former mass enhancement completely and there remains still a large mass enhancement effect. In addition, we find that the mass enhancement factor becomes large with increasing the on-site repulsion $U$ and the density of state (DOS) at the Fermi energy $\rho(0)$. According to many current reseaches, such large $U$ and $\rho(0)$ enhance the effective interaction between quasiparticles, therefore the superconducting transition temperature $T_c$ increases. On the other hand, the large mass enhancement leads the reduction of the energy scale of quasiparticles, as a result, $T_c$ is reduced. When we discuss $T_c$, we have to estimate these two competitive effects.}
 
\kword{mass enhancement, Hubbard model, fourth order, perturbation, selfenergy }

\begin{document}
\maketitle

\section{Introduction} 
In recent decades, superconductivity in strongly correlated electron systems has been intensively investigated, motivated by the discoveries of various superconductors including high-$T_c$ cuprates\cite{bednorz}, other transition metal oxides such as $\mathrm{Sr_2RuO_4}$\cite{maeno}, heavy Fermion compounds\cite{steglich} and molecular conductors\cite{jerome}. In order to describe the systems theoretically, the Hubbard model has been often used as a minimal model including the on-site repulsive term. From the viewpoint of weak coupling, some authors calculated the momentum dependence of the effective interaction using the fluctuation-exchange approximation\cite{bickers}\cite{yanase} and the perturbation theory\cite{yanase}, and then derived the anisotropic pairing state from the on-site repulsive interaction. In such studies, to obtain the realistically observable transition temperature $T_c$, one increases the magnitude of the bare interaction $U$. Large $U$ enhances the effective interaction of quasiparticle, therefore $T_c$ increases. However, it also enhances the effective mass\cite{yamada}, which leads the reduction of the energy scale of quasiparticles, as a result, $T_c$ is reduced. When we discuss $T_c$ quantitatively, we have to estimate these two competitive effects.

In third order perturbation theory (TOPT)\cite{yanase}, the third order terms for selfenergy cancel out completely for the case with the particle-hole symmetry, which occurs in the half-filled case of $t' = 0$. In this case, only one diagram in the second order contributes to the mass enhancement. For the doped case, the third order terms give the negative contribution with large $U$, that is, the effective mass can not become large. To obtain more reliable results, we need to include higher order terms. The $fourth$ order terms for the normal selfenergy have never been studied before in lattice systems. They are composed of two kinds of the particle-hole diagrams, the particle-particle diagram and the vertex corrections. These particle-hole and particle-particle diagrams give the large mass enhancement. The particle-particle diagram has the opposite sign to that in third order. As a result, when we calculated $T_c$ in TOPT, we might have under-estimated the mass enhancement i.e. the reduction of the energy scale. 

 For pairing interaction, Nomura $et\ al$. calculated the fourth order terms\cite{nomura}. They showed the convergence of the pairing interaction is good for moderately strong $U$, when the system is near the half-filling. But they did not include the normal selfenergy. In this paper, we calculate the mass enhancement factor by expanding the normal selfenergy up to the fourth order in the Hubbard model on the two-dimensional square lattice. We consider the cases that the system is near the half-filling. We hope that our quantitative investigation makes clear the essential mechanism determing $T_c$ and explains the differences of observed $T_c$ between the high-$T_c$ systems\cite{zheng}\cite{sasaki}, YBa$_2$Cu$_3$O$_{7-\delta}$ and La$_{2-x}$Sr$_x$CuO$_4$ and so on.

\section{Formulation}
We consider the following single band Hubbard model on the two-dimensional square lattice, 
\begin{align}
H &= H_0 + H_{\mathrm{int}} , \\
H_0 &= \displaystyle \sum_{\vec{k}\sigma} \xi(\vec{k}) c^{\dagger}_{\vec{k}\sigma} c_{\vec{k}\sigma} , \\
H_{\mathrm{int}} &= \frac{U}{2N} \sum_{\vec{k}_i} \sum_{\sigma \neq \sigma'} \delta_{\vec{k}_1+\vec{k}_2 , \vec{k}_3+\vec{k}_4} c^{\dagger}_{\vec{k}_1\sigma} c^{\dagger}_{\vec{k}_2\sigma'} c_{\vec{k}_3\sigma'} c_{\vec{k}_4\sigma} .
\end{align}
From the tight binding approximation, energy dispersion $\xi(\vec{k})$ is given by
\begin{align}
\xi(\vec{k}) = & -2 t \Bigl( \cos {k_x} + \cos {k_y} \Bigr) -4 t' \cos {k_x} \cos {k_y} - \mu .
\end{align} 
Here, $t$ and $t'$ are the nearest- and the next-nearest-neighbor hopping integrals respectively, and $U$ is the on-site repulsion. Hereafter, we fix $t = 1$, and calculate physical quantities by changing the other variables $t'$, $U$, electron filling $n$ and the temperature of the system $T$. The bare Green's function is given as
\begin{align}
G^{(0)}(k) = \frac{1}{i\omega_n - \xi(\vec{k})}.
\end{align}
Here, $k$ is a shorthand notation as $k = (\vec{k},i\omega_n)$ and $\omega_n = (2n+1) \pi T$ (n : integer) is a fermion Matsubara frequency. The chemical potential $\mu$ is determined so as to satisfy the following equation, 
\begin{align}
n = 2 \frac{T}{N}\sum_{k} G^{(0)}(k),
\end{align}
where N is the number of the lattice sites. The bare susceptibility $\chi_0(q)$ is given as
\begin{align}
\chi_0(q) = - \frac{T}{N} \sum_k G^{(0)}(k+q) G^{(0)}(k),
\end{align}
where $q$ is a shorthand notation as $q = (\vec{q},i\Omega_m)$ and $\Omega_m = 2 m \pi T $ (m : integer) is a boson Matsubara frequency.

The normal selfenergy is expanded perturbatively in $U$ as follows,
\begin{align}
\Sigma(k) = \Sigma^{(2)}(k) + \Sigma^{(3)}(k) + \Sigma^{(4)}(k) + \cdots ,
\end{align}
here $\Sigma^{(n)}(k)$ is simply proportional to $U^n$ because there is no momentum dependence in the on-site repulsion $U$. We evaluate the normal selfenergy up to the fourth order with respect to $U$. The diagrammatic and analytic expressions are given in Appendix A.

In this paper, we mainly discuss the mass enhancement factor $z^{-1}(\vec{k})$, which is given as 
\begin{align}
z^{-1}(\vec{k}) = 1 - \frac{\partial \Sigma_{\mathrm{R}}(\vec{k},\omega)}{\partial \omega} \Big|_{\omega \to 0}.
\end{align}
Here, $\Sigma_{\mathrm{R}}(\vec{k},\omega)$ is real frequency retarded selfenergy, which is obtaind by performing the analytic continuation of $\Sigma(\vec{k},i\omega_n)$ from the upper half plane with use of Pad$\acute{\mathrm{e}}$ approximation.

We take $64 \times 64$ $\vec{k}$-meshs for the first Brillouin zone and 512 Matsubara frequencies in the numerical calculations. We fix the temperature of the system $T = 0.01$. The other details of the calculation are given in Appendix A. 

\section{Results}
We consider the cases that the system is near the half-filling. These are similar situations to high-$T_c$ cuprates. We fixed the parameters $t = 1$, $t' = -0.15$ and $T = 0.01$. In Fig.1, we show the Fermi surfaces for $n = 0.9$ and $n = 1.1$. As we set the half-filling $n = 1$, the electron filling $n = 0.9$ ($n = 1.1$) represents the hole-doped (electron-doped) case. We can see the ($\pi, \pi$) nesting, which yeilds the strong antiferromagnetic (AF) fluctuations. The band width of these dispersions equals 8.0. Figure 2 shows the density of state in these two cases. The DOS at the Fermi level for $n = 0.9$ is larger, bacause when we introduce the negative $t'$ at the half-filling, the level of the van Hove singularity falls down. We can see the strong peak of AF fluctuations at ($\pi, \pi$) in Fig.3. The peak is stronger in the hole-doped case because of the strong nesting property and the large $\rho(0)$.

\begin{figure}[t]
\begin{center}
\includegraphics[width=6cm]{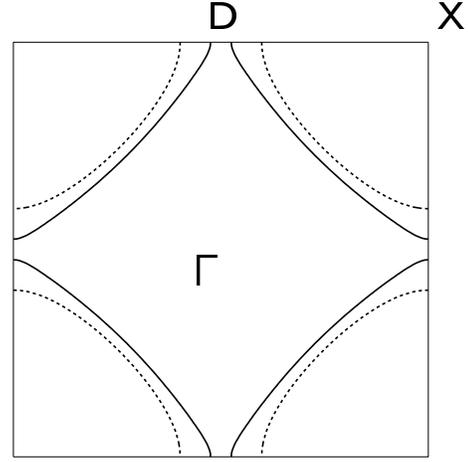}
\end{center}
\caption{The Fermi surfaces in the case of $t = 1$, $t' = -0.15$ and $T = 0.01$. The solid (dashed) line is the result for $n = 0.9$ ($n = 1.1$). The symbols $\mathrm{\Gamma}$, D and X represent the symmetry points (0,0), (0,$\pi$) and ($\pi$, $\pi$) respectively.}
\label{f1}
\end{figure}

\begin{figure}[t]
\begin{center}
\includegraphics[width=8cm]{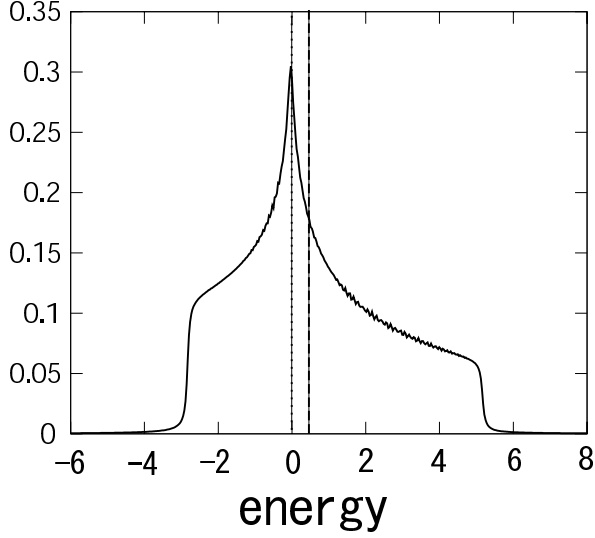}
\end{center}
\caption{The density of state in the case of $t = 1$, $t' = -0.15$ and $T = 0.01$. The dotted line and the zero energy represent the Fermi level for $n = 0.9$. And the dashed line represents the Fermi level for $n = 1.1$.}
\label{f2}
\end{figure}

\begin{figure}[t]
\begin{center}
\includegraphics[width=8cm]{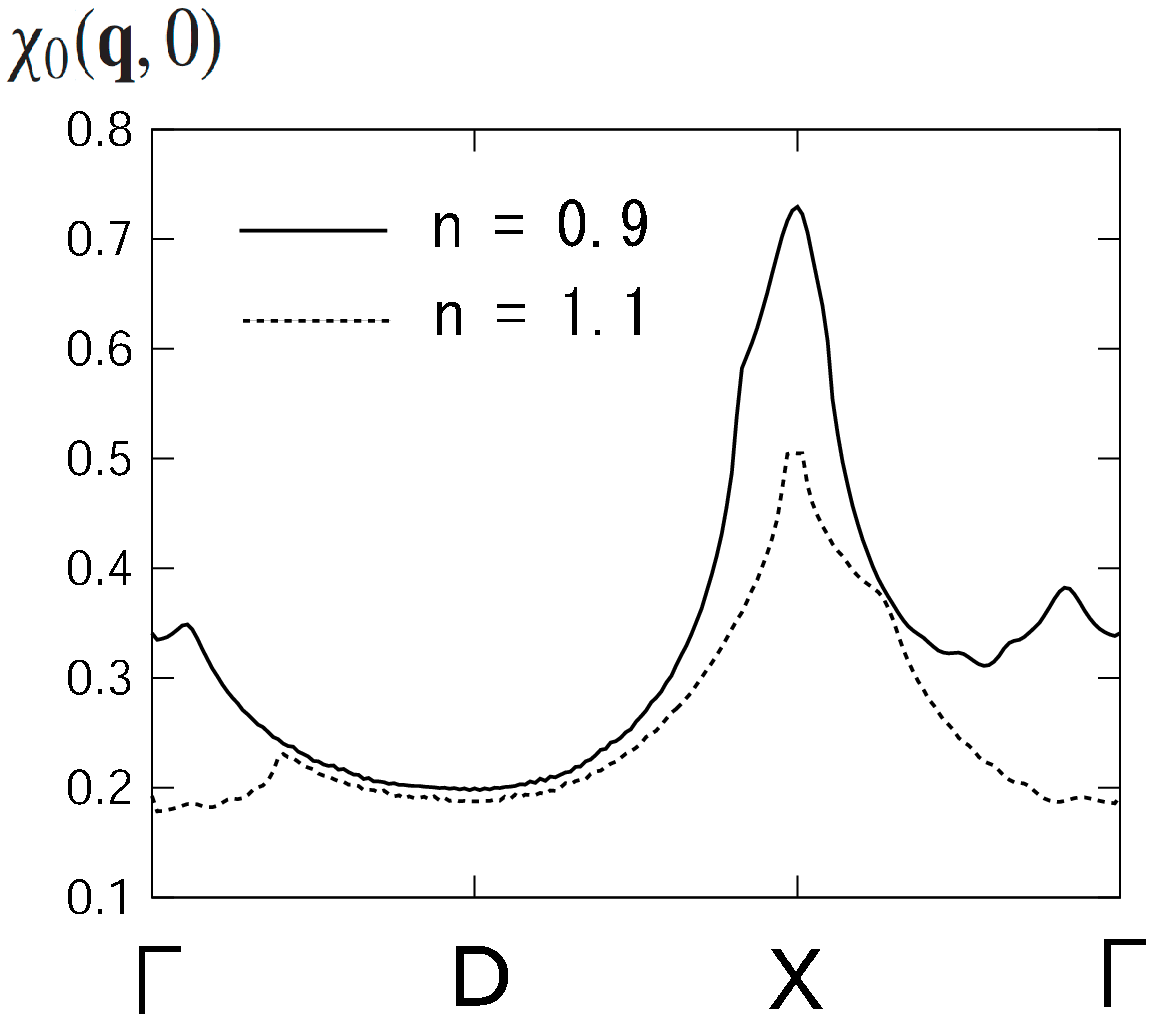}
\end{center}
\caption{The bare spin susceptibilities $\chi_0(\vec{q},0)$ in the case of $t = 1$, $t' = -0.15$ and $T = 0.01$. The solid (dashed) line is the result for $n = 0.9$ ($n = 1.1$).}
\label{f3}
\end{figure}

We have calculated the mass enhancement factor $z^{-1}(\vec{k})$ in those two cases $n = 0.9$ and $n = 1.1$. First, we show the result for the hole-doped case $n = 0.9$ in Fig.4. By introducing the fourth order terms, we obtain the large effective mass at the Fermi surface, which is almost twice as large as that calculated in TOPT. This result solves the problem of the reduction of $z^{-1}(\vec{k})$ due to the third order terms as we mentioned before. We can define the value $z^{-1}$ as the average of $z^{-1}(\vec{k})$ at the Fermi point near $(0, \pi)$ and that near $(\pi/2, \pi/2)$. In Fig.5, we show the $U$-dependence of $z^{-1}$ in the case of $n = 0.9$. With increasing $U$, the mass enhancement factor increases. From this calculation, we can consider $z^{-1}$ to be a function of $U$ and estimate the coefficients of $U^n$, such as 
\begin{align}
z^{-1} = 1 + 0.0564 U^2 - 0.0048 U^3 + 0.0063 U^4.
\end{align}

Next, we show the result for $n = 1.1$, in Fig.6. We can also see the large mass enhancement, but it is relatively low. This result is related to the fact that $\rho(0)$ is lower than that in the case of $n = 0.9$. In Fig.7, we show the $U$-dependence of $z^{-1}$ and estimate $z^{-1}$ to be 
\begin{align} 
z^{-1} = 1 + 0.0273 U^2 - 0.0011 U^3 + 0.0014 U^4
\end{align}
When we set a large $U$ value, we can also obtain the large mass enhancement.

As we discuss in Appendix B, roughly speaking, the contribution from all the fourth order terms is about half as much as the one from the diagram (4A) as the result of the cancellation in the hole-doped case ($n = 0.9$). In the electron-dope case ($n = 1.1$), the cancellation is more remarkable, and the ratio of the coefficient of $U^4$ to that of $U^2$ in eq.(11) is smaller than that in eq.(10).

\begin{figure}[t]
\begin{center}
\includegraphics[width=8cm]{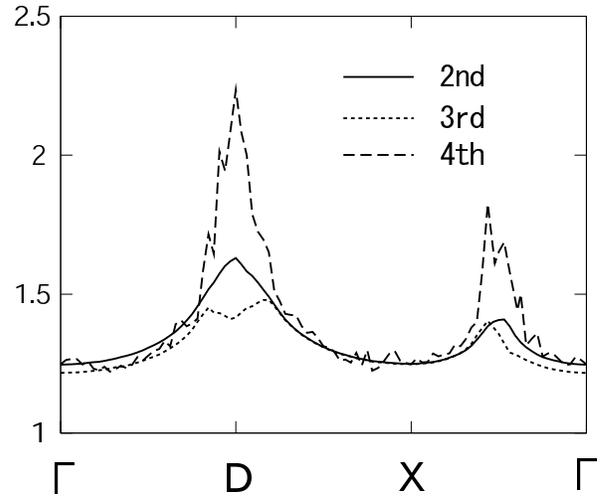}
\end{center}
\caption{The mass enhancement factor $z^{-1}(\vec{k})$ in the case of $n = 0.9$ and $U = 3$. The solid, dotted and dashed lines are the results in the calculations up to the second, third and fourth order, respectively. $z^{-1}(\vec{k}) = 1$ means the mass of the free electrons.}
\label{f4}
\end{figure}

\begin{figure}[t]
\begin{center}
\includegraphics[width=8cm]{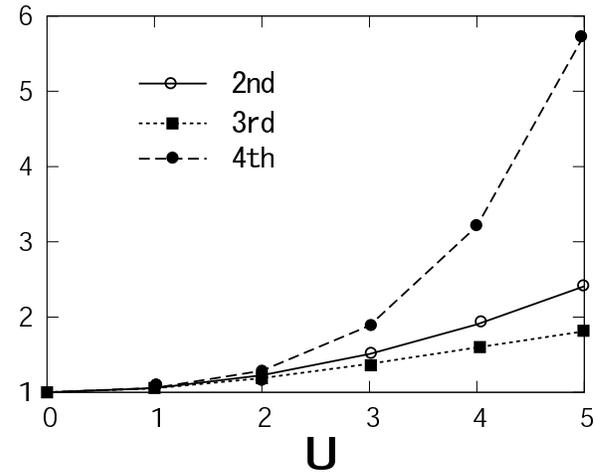}
\end{center}
\caption{The $U$-dependence of the mass enhancement factor $z^{-1}$ in the case of $n = 0.9$. The solid, dotted and dashed lines are the results in the calculations up to the second, third and fourth order, respectively. $z^{-1}$ is the average of $z^{-1}(\vec{k})$ at the Fermi point near $(0, \pi)$ and that near $(\pi/2, \pi/2)$.}
\label{f5}
\end{figure}

\begin{figure}[t]
\begin{center}
\includegraphics[width=8cm]{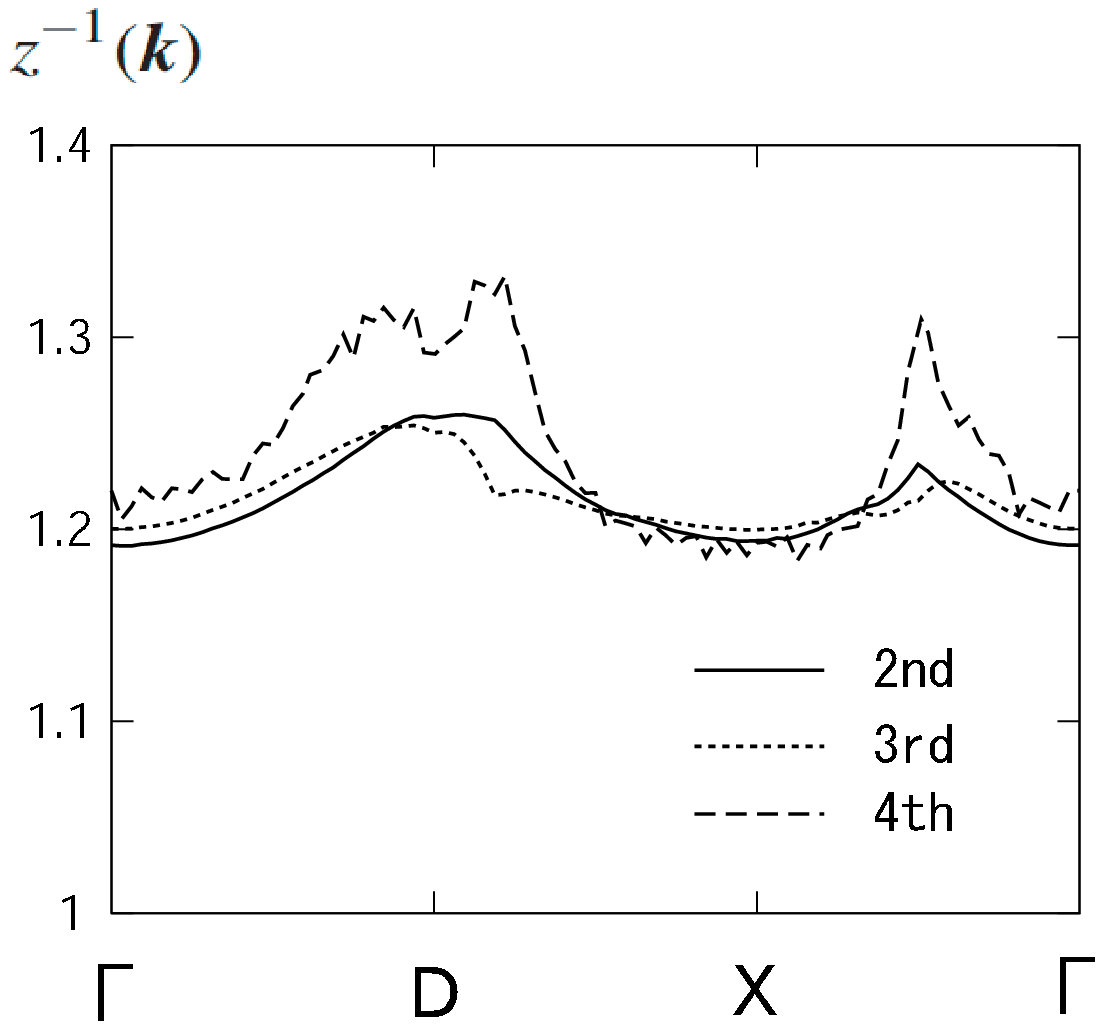}
\end{center}
\caption{The mass enhancement factor $z^{-1}(\vec{k})$ in the case of $n = 1.1$ and $U = 3$. The solid, dotted and dashed lines are the results in the calculations up to the second, third and fourth order, respectively. $z^{-1}(\vec{k}) = 1$ means the mass of the free electrons.}
\label{f6}
\end{figure}

\begin{figure}[t]
\begin{center}
\includegraphics[width=8cm]{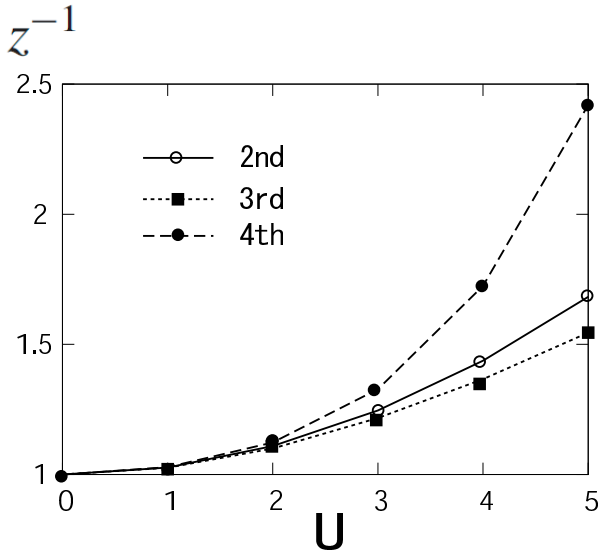}
\end{center}
\caption{The $U$-dependence of the mass enhancement factor $z^{-1}$ in the case of $n = 1.1$. The solid, dotted and dashed lines are the results in the calculations up to the second, third and fourth order, respectively.}
\label{f7}
\end{figure}

\section{Discussion}
In this paper, we have investigated the the mass enhancement factor perturbatively up to the fourth order in the Hubbard model. We obtained the large mass enhancement on the Fermi surface compared with the result in TOPT, and it becomes large with increasing $U$ and $\rho(0)$. 

 In this section, we discuss the relation between mass renormalization and $T_c$. Some authers have described the Fermi liquid theory for the strongly correlated electron systems\cite{yamada}\cite{nisikawa}. Based on this idea, electrons form the quasiparticles with effective mass and finite lifetime due to the effect of the interaction U. Therefore, we have to start with the estimation of the renormalization factor $z$, then calculate the momentum dependence of the effective interaction between the renormalized quasiparticles. In that scheme, the renormalized parameters are given as $t^{(\mathrm{r})} = z t$, ${t'}^{(\mathrm{r})} = z t'$, $U^{(\mathrm{r})} = z U$, where the index (r) denotes renormalization due to the normal selfenergy correction and z is the renormalization factor. Therefore, we consider that the transition temperature in the renormalized scheme $T^{(\mathrm{r})}_c$ is given as
\begin{align}
T^{(\mathrm{r})}_c = z T_c 
\end{align}
This relation show the reduction of $T_c$ due to small $z$ as we mentioned before. Taking our calculation into account, this tendency is considered to be more marked by introducing the fourth order terms. We have to perform the calculation of $T_c$ based on the fourth order selfenergy correction. This is further investigation in the future.

Here we would like to stress the following point. The anisotropic part of quasiparticle interaction determines the superconductivity. On the other hand, the mass enhancement factor is determined by total interaction including isotropic interaction. Therefore, in principle, we have independent two parameters which determine the transition temperature. The two parameters are determined by details of electronic structures, depending on the Fermi surface and $d$ or $f$ electron numbers.

\section{Conclusions}
We investigate the mass enhancement factor by expanding the normal selfenergy perturbatively up to the fourth order with respect to $U$ in the Hubbard model. We consider the cases that the system is near the half-filling on the two-dimensional square lattice. These are similar situations to high-$T_c$ cuprates. As results of the calculations, by introducing the fourth order terms, we obtain the large mass enhancement on the Fermi surface compared with the result in TOPT. This is mainly due to the fourth order particle-hole and particle-particle terms. Although the other forth order terms have effect of reducing the effective mass, this effect does not cancel out the former mass enhancement completely and there remains still a large mass enhancement effect. In addition, we find that the mass enhancement factor becomes large with increasing the on-site repulsion $U$ and the density of state at the Fermi energy $\rho(0)$. According to many current reseaches, such large $U$ and $\rho(0)$ enhance the effective interaction between quasiparticles, therefore $T_c$ increases. On the other hand, the large mass enhancement leads the reduction of the energy scale of quasiparticles, as a result, $T_c$ is reduced. When we discuss $T_c$, we have to estimate these two competitive effects. This is the important unified theory for all the strongly correlated electron systems from cuprates to heavy fermion compounds.

\section*{Acknowledgements}
Numerical calculation in this work was carried out at the Yukawa Institute Computer Facility.

\appendix 
\section{Details of Calculation}
The normal selfenergy is expanded perturbatively in $U$ as follows,
\begin{align}
\Sigma(k) = \Sigma^{(2)}(k) + \Sigma^{(3)}(k) + \Sigma^{(4)}(k) + \cdots ,
\end{align}
here $\Sigma^{(n)}(k)$ is simply proportional to $U^n$. These perturbation terms are represented by using the bare Green's function $G^{(0)}(k)$, and the bare susceptibilities,
\begin{align}
\chi_0(q) &= - \frac{T}{N} \sum_k G^{(0)}(k+q) G^{(0)}(k)\\
\phi_0(q) &= - \frac{T}{N} \sum_k G^{(0)}(q-k) G^{(0)}(k).
\end{align}
The diagrams for the normal selfenergy up to the fourth order are shown in Fig.A$\cdot$1, and the analytic representations for them are given in the followings, where $k$ is the shorthand notation of the external momentum and frequency, and $p_i$'s ($q_i$'s) are those of the internal momenta and frequencies for fermions (bosons). 

\begin{figure}[t]
\begin{center}
\includegraphics[width=8cm]{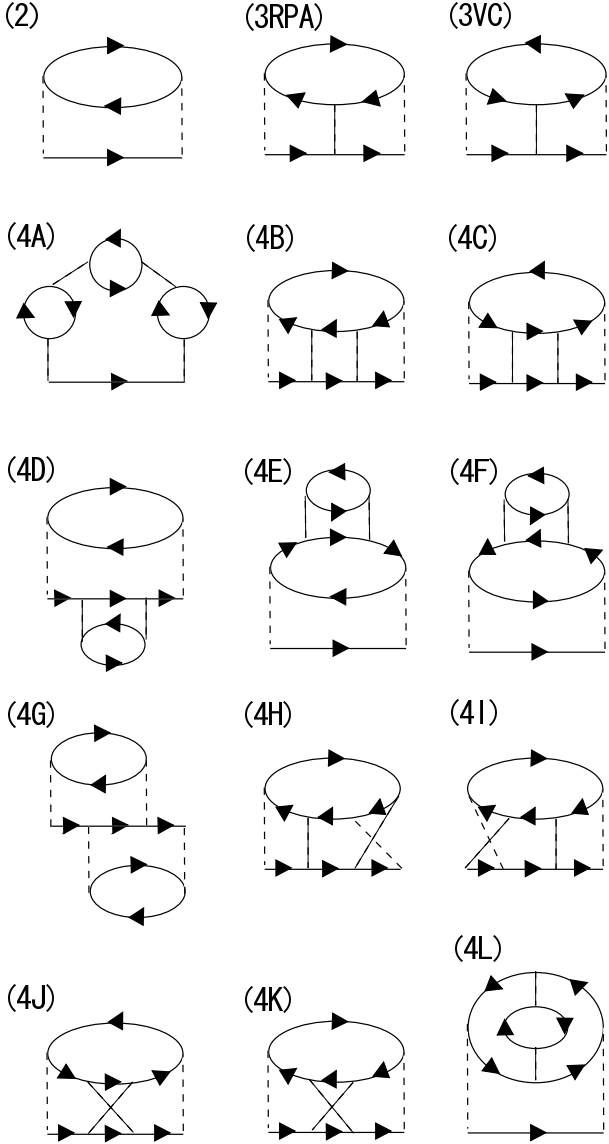}
\end{center}
\caption{The Feynman diagrams of the normal selfenergy up to the fourth order. Solid and dashed lines correspond to the bare Green's function and the interaction, respectively.}
\label{fA1}
\end{figure}

\subsection*{The second order term }
\begin{align}
\Sigma^{(2)}(k) = U^2 \frac{T}{N} \sum_q\ \chi_0(q)\ G^{(0)}(k-q).
\end{align}
\subsection*{The third order term }
\begin{align}
\Sigma^{(3)}(k) = \Sigma^{(3 \mathrm{RPA})}(k) + \Sigma^{(3 \mathrm{VC})}(k),
\end{align}
where
\begin{align}
\Sigma^{(3 \mathrm{RPA})}(k) &= U^3 \frac{T}{N} \sum_q\ \chi_0(q)^2\ G^{(0)}(k-q),\\ 
\Sigma^{(3 \mathrm{VC})}(k) &= U^3 \frac{T}{N} \sum_q\ \phi_0(q)^2\ G^{(0)}(q-k).
\end{align}
\subsection*{The fourth order term }
\begin{align}
\Sigma^{(4)}(k) = \Sigma^{(4 \mathrm{A})}(k) + \Sigma^{(4 \mathrm{B})}(k) + \cdots + \Sigma^{(4 \mathrm{L})}(k), 
\end{align}
where 
\begin{align}
\Sigma^{(4 \mathrm{A})}(k) =\ & U^4 \frac{T}{N} \sum_q\ \chi_0(q)^3\ G^{(0)}(k-q), \\
\Sigma^{(4 \mathrm{B})}(k) =\ & U^4 \frac{T}{N} \sum_q\ \chi_0(q)^3\ G^{(0)}(k-q), \\
\Sigma^{(4 \mathrm{C})}(k) =\ & U^4 \frac{T}{N} \sum_q\ \phi_0(q)^3\ G^{(0)}(q-k), \\
\Sigma^{(4 \mathrm{D})}(k) =\ & U^4 \frac{T}{N} \sum_p\ \chi_0(k-p)\ G^{(0)}(p)^2\ \Sigma^{(2)}(p), \\
\Sigma^{(4 \mathrm{E})}(k) =\ & U^4 \frac{T}{N} \sum_p\ \chi_0(p-k)\ G^{(0)}(p)^2\ \Sigma^{(2)}(p), \\
\Sigma^{(4 \mathrm{F})}(k) =\ & U^4 \frac{T}{N} \sum_p\ \phi_0(k+p)\ G^{(0)}(p)^2\ \Sigma^{(2)}(p), \\
\Sigma^{(4 \mathrm{G})}(k) =\ & U^4 \big( \frac{T}{N} \big)^2 \sum_{q_1} \sum_{q_2}\ \chi_0(q_1)\ \chi_0(q_2) \nonumber \\
&\times G^{(0)}(k-q_1)\ G^{(0)}(k-q_2) \nonumber \\
&\times G^{(0)}(k-q_1-q_2), \\
\Sigma^{(4 \mathrm{H})}(k) =\ & -\ U^4 \big( \frac{T}{N} \big)^2 \sum_{q_1} \sum_{q_2}\ \chi_0(q_1)\ \phi_0(q_2) \nonumber \\
&\times G^{(0)}(k-q_1)\ G^{(0)}(q_2-k) \nonumber \\
&\times G^{(0)}(q_1+q_2-k), \\
\Sigma^{(4 \mathrm{I})}(k) =\ & -\ U^4 \big( \frac{T}{N} \big)^2 \sum_{q_1} \sum_{q_2}\ \chi_0(q_1)\ \phi_0(q_2) \nonumber \\
&\times G^{(0)}(k-q_1)\ G^{(0)}(q_2-k) \nonumber \\
&\times G^{(0)}(q_1+q_2-k), \\
\Sigma^{(4 \mathrm{J})}(k) =\ & U^4 \big( \frac{T}{N} \big)^3 \sum_q \sum_{p_1} \sum_{p_2}\ \chi_0(q-p_1-p_2) \nonumber \\
&\times G^{(0)}(p_1)\ G^{(0)}(p_2)\ G^{(0)}(q-p_1) \nonumber \\
&\times G^{(0)}(q-p_2)\ G^{(0)}(q-k), \\
\Sigma^{(4 \mathrm{K})}(k) =\ & U^4 \big( \frac{T}{N} \big)^3 \sum_q \sum_{p_1} \sum_{p_2}\ \phi_0(q+p_1+p_2) \nonumber \\
&\times G^{(0)}(p_1)\ G^{(0)}(p_2)\ G^{(0)}(q+p_1) \nonumber \\
&\times G^{(0)}(q+p_2)\ G^{(0)}(k-q), \\
\Sigma^{(4 \mathrm{L})}(k) =\ & -\ U^4 \big( \frac{T}{N} \big)^3 \sum_q \sum_{p_1} \sum_{p_2}\ \chi_0(p_1-p_2) \nonumber \\
&\times G^{(0)}(p_1)\ G^{(0)}(p_2)\ G^{(0)}(q+p_1) \nonumber \\
&\times G^{(0)}(q+p_2)\ G^{(0)}(k-q). 
\end{align}
\\

We take $64 \times 64$ $\vec{k}$-meshs for the first Brillouin zone and 512 Matsubara frecencies in the numerical calculations. We evaluate the second order term, the third order terms, the fourth order (4A), (4B), $\cdots$ , (4E) and (4F) terms by using the fast Fourier transformation (FFT) algorithm because the summations included in them all have convolution forms. In the cases of the fourth order (4G), (4H) and (4I) terms, we fix the external momentum and frequency $k$. Then, the full summations with respect to the internal momenta and frequencies are performed by using the FFT algorithm. 

The evaluation of the fourth order (4J), (4K) and (4L) diagrams is very difficult. The summations of momenta and frequencies for them are not reduced to convolution forms even if we fix the external momenta $k$, so the estimated computation time is not realistic. Therefore, we take the following approximation. First, we notice that these diagrams have such forms as only one bare Green's function contains the enternal momenta $k$, then we separate it and write the rest as $A(q)$. For example, in the case of (4J), 
\begin{align}
\Sigma^{(4 \mathrm{J})}(k) \equiv&\ U^4 \frac{T}{N} \sum_q A(q)\ G^{(0)}(q-k) \\
A(q) =&\ \big( \frac{T}{N} \big)^2 \sum_{p_1} \sum_{p_2}\ \chi_0(q-p_1-p_2) \nonumber \\
&\times G^{(0)}(p_1)\ G^{(0)}(p_2)\ G^{(0)}(q-p_1) \nonumber \\
&\times G^{(0)}(q-p_2). 
\end{align}
The equation (A$\cdot$21) has convolution form, therefore the remaining problem is the calculation of $A(q)$. We perform the Fourier transformation and consider the following quantity,
\begin{align}
A(r) =& \sum_{q}\ A(q)\ e^{iqr}.
\end{align}
This quantity $A(r)$ is reduced to  
\begin{align}
A(r) =& \sum_{q'}\ \chi_0(q')\ e^{iq'r} \nonumber \\
& \times \frac{T}{N} \sum_{p_1}\ G^{(0)}(p_1)\ G^{(0)}(q'+p_1)\ e^{ip_1r} \nonumber \\
& \times \frac{T}{N} \sum_{p_2}\ G^{(0)}(p_2)\ G^{(0)}(q'+p_2)\ e^{ip_2r},
\end{align}
where $r$ is the shorthand notation $r = (\vec{r}, \tau)$, and we have defined the variable $q' \equiv q - p_1 - p_2$. We can see that the summations in eq.(A$\cdot$24) can be performed with use of the FFT algorithm for fixed $r$. We calculate $A(r)$ for only $9 \times 9$ $\vec{r}$-region around $|\vec{r}| = 0$, because with increasing $|\vec{r}|$, A(r) converges rapidly. This is a similar way which Nomura $et\ al.$ estimate the non-Parquet 'envelope' diagram\cite{nomura}. This cut-off of the large $|\vec{r}|$ components means that in $\vec{q}$-space the small structure of $A(q)$ is neglected. Here, we note that although the shape of $A(q)$ becomes rather smooth, this approximation hardly affect the normal selfenergy quantitatively. As for calculating in $\tau$-space, we reduce the $\tau$-mesh to 256, then we fill the middle points with the averages. We have verified that this reduction also has no quantitative problem. Finally, we can obtain $\Sigma^{(4 \mathrm{J})}(k)$ by calculating the convolution of $A(q)$ and $G^{(0)}(q-k)$ with use of the FFT algorithm.

\section{Contribution from Each Diagram in Fourth Order}
In this section, we briefly discuss the contribution from each diagram in fourth order. The RPA diagrams (4A) and (4B) give the large mass enhancement, and so is the particle-particle diagram (4C). Only these three diagrams enhance the effective mass, and then the other nine diagrams (4D), $\cdots$, (4K) and (4L) reduce it. If we neglect the latters, we obtain about six times larger coefficient of $U^4$ in eq.(10), that is too overestimating. The diagrams (4G), (4H) and (4I) give large negative contribution to the mass enhancement, which mainly cancel the formers' positive contribution however not completely. On the other hand, the one from the diagrams (4D), (4E) and (4F) is very small. And the fact does not change even if we collect this type of diagram up to infinite order. In the case of the diagram (4J), (4K) and (4L), their contribution is about half as much as that from the diagram (4G). This is considered to be related to the fact that there is one extra summation which can not be reduced to $\chi_0$ or $\phi_0$. As the result of the cancellation, roughly speaking, the contribution from all the fourth order terms is about half as much as the one from the diagram (4A) in the hole-doped case ($n = 0.9$). In the electron-dope case ($n = 1.1$), the cancellation is more remarkable, and the ratio of the coefficient of $U^4$ to that of $U^2$ in eq.(11) is smaller than that in eq.(10).

\end{document}